# MESH

**Radu Vultur, student**
Faculty of Computers and Applied Computer Science,
"Tibiscus" University, Timișoara, România

ABSTRACT. Whether you just want to take a peek of a remote computer status, or you want to install the latest version of a software on several workstations, you can do all of this from your computer. The networks are growing, the time spent administering the workstations increases and the number of repetitive tasks is going sky high. But here comes **MESH** to take that load off your shoulders. And because of SMS commands you can take this "command center" wherever you will go. Just connect a GSM phone to the computer (using a cable, IrDA or Bluetooth) and lock/restart/shutdown computers from your LAN with the push of a cell phone button. You can even create your own SMS commands.
This is **MESH** – the network administrator's Swiss knife.

## 1. MESH - technical information

This application is built on Microsoft Dot Net framework, using C# and Visual Basic. The software was designed to support a plug-in system (Winamp style) in order to make further modifications or adding new features a breeze. Each module (or plug-in) is kept in a separate DLL file located in Plugging folder.

**MESH** makes extensive use of WMI (Windows Management Instrumentation).

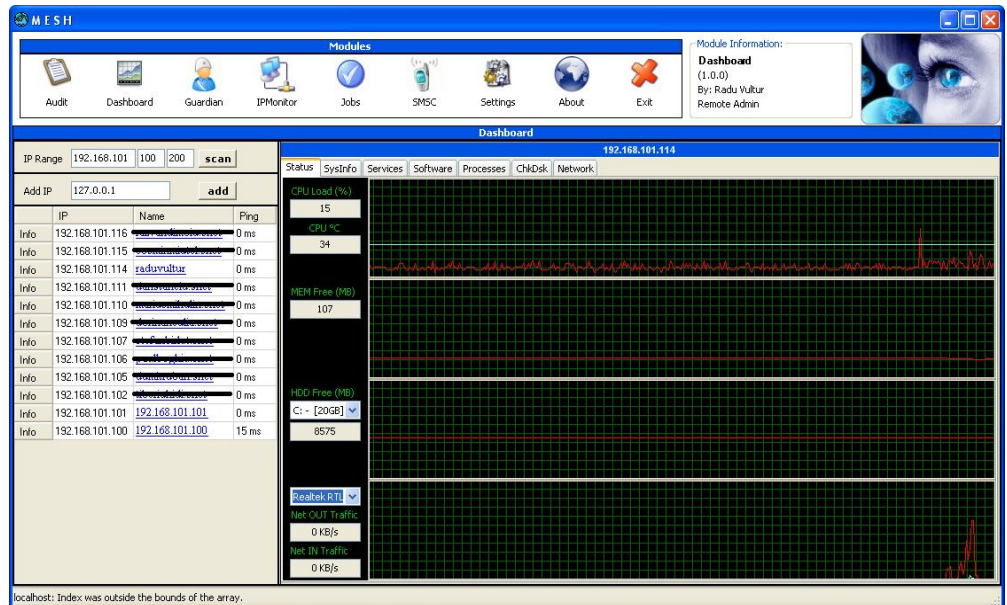

***Fig.1:*** *MESH - the Dashboard plug-in*

The current list of modules enumerates:
- Audit - creates reports with installed software and hardware configurations
- Dashboard - monitor a specific computer and run various remote tasks on it
- Guardian - exports the status of AntiVirus and Firewall protection
- IPMonitor - monitors a computer online status and services availability
- Jobs - creates and runs batch jobs on multiple computers
- Killer - doesn't allow a specific process to run on selected computer
- SMSC - listens for SMS commands and runs them when received

## 2. Modules in detail

Audit

Creates reports with installed software and hardware configurations. This report is stored in an XML file but you can choose to export it in Html format also. This is very useful when creating the whole company software and/or hardware audit. You don't have to wait for hours to get each computer report and then to centralize all the data. Use *Audit* and you have all this information's with a single mouse click.

Dashboard

With *Dashboard* you can monitor a specific computer and run various remote tasks on it. Just select the desired computer from the list and watch the CPU/Memory/Hard disk usage and the network card traffic. You have also the possibility to kill a process or create a new one, stop or start a service, install or uninstall software applications and even run Chides command from the comfort of your own desk, without having to actually walk to that computer. Shutdown or Restart that computer is also a "one-click" action offered by *Dashboard*.

Guardian

Reports the status of AntiVirus and Firewall protection status of the selected computers. No actions here (for security reasons), you just can see the status information which includes: product name and version, if it's up to date (virus signatures updated) and if scanning is enabled. For firewalls the information consists of: company, product, and version and if it's enabled.

IPMonitor

Monitors a computer online status and services availability. If the computer check fails you can choose among various actions like: no action, start a process/play a file or send an email. It's good to know that the email will contain the trace route information for that IP.

Jobs

Creates and runs batch jobs on multiple computers. After you have selected the computers you want this job to run you can select the action and action parameters. The current supported actions are: install/uninstall software, start/kill a process, start/stop a service, reboot, and shutdown and send a message.

Killer

Doesn't allow a specific process to run on selected computer. You just have to type the process name, select the computer, add this rule to the list and start *Killer*.

SMSC

This is short for SMS Commander. It will check a connected GSM phone for SMS and when a command SMS is found it checks to see if the phone number that has sent the message is allowed to run commands and if it is, then the command is run. **MESH** has some built-in commands to lock / reboot / shutdown the computer and to start or kill a process. Beside these, you can create your own custom commands.

## 3. Example of SMS communication

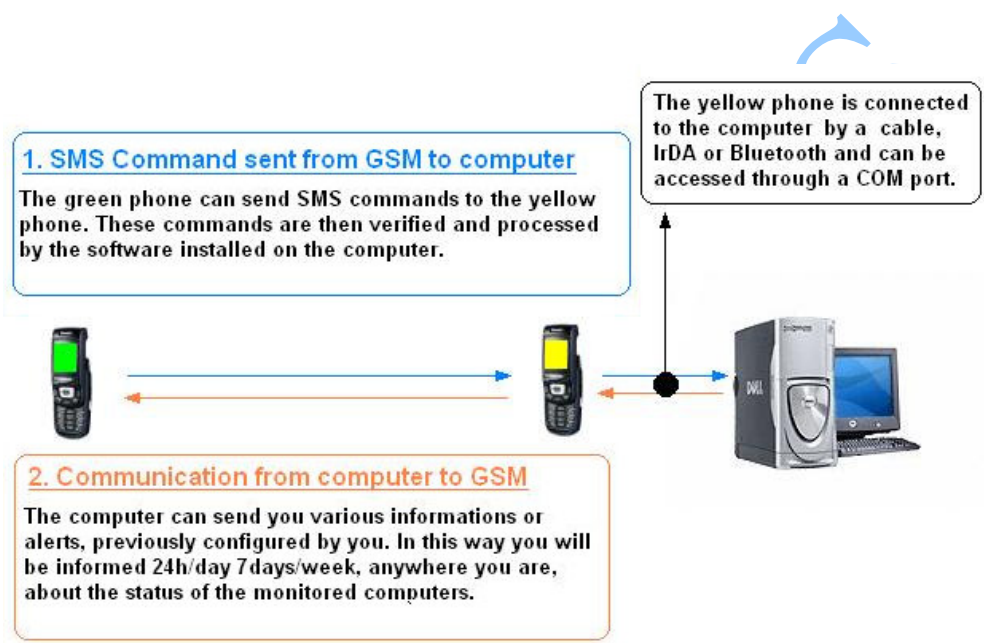

*Fig. 2: Two way communication*

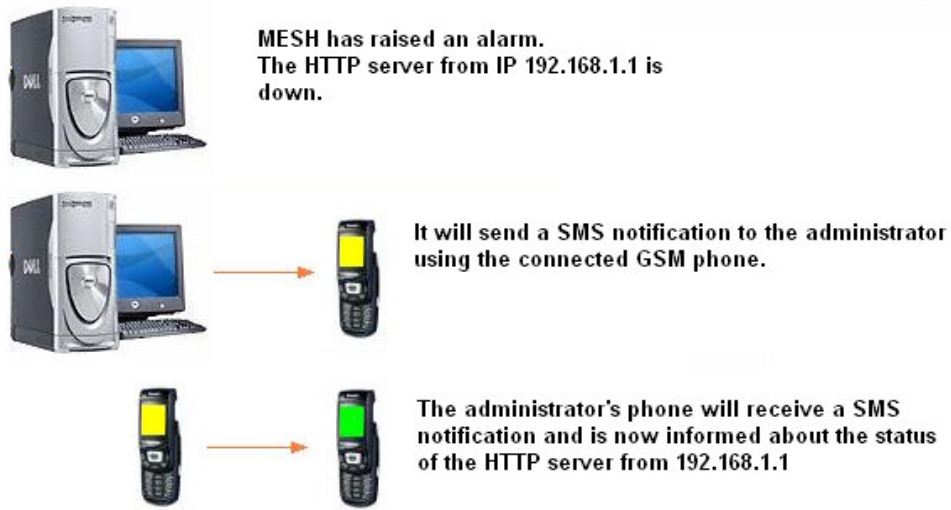

*Fig. 3:* Scenario 1

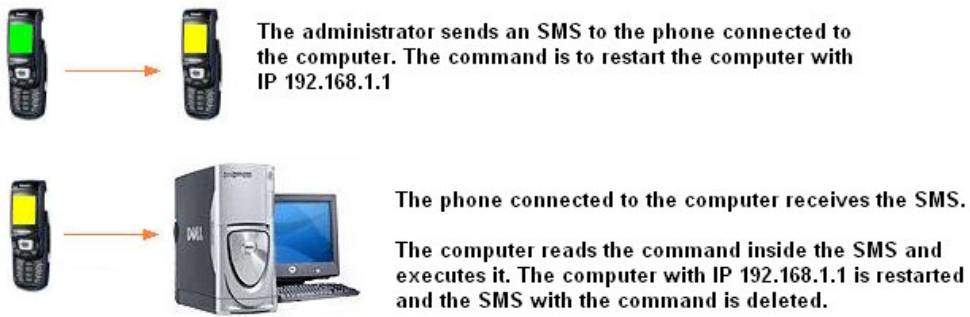

*Fig. 4:* Scenario 2

**Conclusions**

At startup, **MESH** is scanning this folder for plugging and adds them in the modules list. The second operation done by the application when starting is to search for the Settings folder and apply all the saved configurations found

265

(for example the list of computers, the username and password which is encrypted).